\documentstyle[aps,preprint]{revtex}
\begin{document}
\title{Heterodyne Detection of Gravitational Waves in a
Michelson Interferometer with a Vibrating Mirror}
\author{ Y.Ben-Aryeh \and Physics Department, Technion , Haifa 32000, Israel}
\date{July 4, 1999}
\maketitle
\begin{abstract}A new heterodyne detection
method is suggested for detecting gravitational waves in a Michelson Interferometer.
The method is based on interference between phase changes which are induced by a vibrating mirror with phase changes which
are due to the gravitational waves. The advantage of using a second order correlation function in the present analysis is discussed.
\end{abstract}

In the present article
a special heterodyne scheme is suggested for detecting a periodic gravitational wave$^{1-4}$, which has a very 
low frequency (relative to optical frequencies) in a Michelson interferometer which has two
long arms (in the order of few kilometers length). Since a polarized gravitational wave is expected to produce very small 
phase differences between the two arms of the interferometer , detection of
such small phase differences$^{1-19}$ is related to optical problems. In the Michelson interferometer$^{20,21}\ $ monochromatic light from a laser with a certain 
wavelength $\lambda $ is split into two
orthogonal beams via a beam splitter. Each beam traverses a path length L and is reflected by a mirror to the beam splitter. Here the beams are recombined and the output is detected by a photodiode. 
The output observed will depend on the relative path difference between the two arms. In the conventional approach , the effective path difference is
 expected to be changed by the gravitational wave which 
will be observed as a shift in the interference pattern. 

In a previous study$^{22}\,$ it has been suggested to insert in one arm of the interferometer a slab of crystal (e.g. KDP) in which the index 
of refraction will be changed
periodically by electrooptic effect$^{21,23}$ , and the phase changes due to the modulation of the optical path difference will interfere with those of the gravitational waves. This idea is however
not feasible since the insertion of the crystal in the interferometer will introduce extra noise terms, and also the crystal will be destroyed by the strong laser beams 
which are used for gravitational waves
detection. Since the mirrors of interferometric detectors of gravitational waves are often suspended by the use of magnetic forces , it is relatively easier to use a periodic magnetic (or electrostatic)
field which would lead to periodic vibrations of one mirror$^{14,19}$. The modulation of the path difference between the two arms of the interferometer due to the vibration of the mirror can lead to
interference with a periodic gravitational wave, which can be observed by heterodyne detection methods. The advantage of using a heterodyne detection method follows from the fact that it is phase
dependent and can concentrate the measurement into a very narrow bandwidth. Since the main noise sources in gravitational waves detectors (e.g. shot noise, thermal noise etc.) are proportional to
the bandwidth of measurement , the narrowing of the bandwidth by the heterodyne detection leads , effectively , to reduction of the noise terms.

The interference of light from the two arms of the
interferometer can be described schematically as
\begin{equation}
\left\langle \in ^{2}\right\rangle \simeq \in _{0}^{2}\left[ 1+\left\langle\cos \left( \Delta \Phi _{1}-\Delta \Phi _{2}\right) \right\rangle \right]
\end{equation}
Here $\Delta \Phi _{1}$ and $\Delta \Phi _{2}$ include all the phase terms in the first and second arms of the interferometer , respectively. $\left\langle {}\right\rangle $
indicates averaged values over times which are very long ,
relative to the time period of the optical wave , which has a frequency $\omega $ . $\in $ is the amplitude of the light output from the interferometer and 
$\in _{0}$ is the the light amplitude in each of
the two arms of the interferometer. We assume that $\left( \Delta \Phi _{1}-\Delta\Phi _{2}\right) $ includes the gravitational phase modulation
\begin{equation}
\Delta \Phi _{grav}=a_{grav}\cos \left(\omega _{g}t\right)
\end{equation}
and phase modulation by the periodic vibration of the mirror
\begin{equation}
\Delta \Phi _{vib}=a_{vib}\cos (\omega _{vib}t)
\end{equation}
By substituting Eqs. (2-3) into
Eq. (1) , developing $\cos \left( \Delta\Phi _{1}-\Delta \Phi _{2}\right) $ with the use of Bessel functions of the first kind$^{23-24}$ , we find that the light 
intensity of the laser output includes the
heterodyne term:
\begin{equation}
I_{Het}\propto \, J_{1}\left( a_{grav}\right) \sin \left( \omega_{g}t\right) 
\times J_{1}\left( a_{vib}\right) \sin \left(\omega_{vib}t\right)
\end{equation}
\newline where $J_{1}$ is the Bessel function of the first kind. By a straight forward calculation of Eq. (4) we find that the output of the interferometer will include
light modulated with the frequencies $\omega _{g}\pm \omega _{vib}$.
% with intensity which is proportional to $J_{ 1}\left( a_{grav}\right) \timesJ_{1}\left( a_{vib}\right) $ . 
For $J_{1}\left(a_{grav}\right) $ one can use the approximation:
\begin{equation}
J_{1}\left( a_{grav}\right) \simeq \,\frac{1}{2}a_{grav}=\left( \frac{\Delta L_{g}}{\lambda }\right) \pi
\end{equation}
Here $\Delta L_{g}$
is the maximal path difference between the two arms of the interferometer , induced by the periodic gravitational wave. For a strong periodic gravitational wave , an order of magnitude
$\frac{\Delta L_{g}}{L}\simeq 10^{-22}$ can be estimated$^{1-19}$ for a path difference L. For 6 kilometers path difference , between the two arms of the interferometer ,  
and $\lambda =10^{-4}cm$ we get
$a_{grav}\simeq 2\times10^{-12}$ which is a very small number. However $J_{1}\left( a_{vib}\right) $ can
 be , relatively , a quite large number. Following the above analysis , we find that the intensity of the light output from the Michelson interferometer can have the
component
\begin{equation}
I_{Het}\propto a_{grav}J_{1}\left( a_{vib}\right) \cos \left[ \left(\omega _{g}+\omega _{vib}\right) t\right]
\end{equation}
In the common method of measuring periodic
gravitational waves by observing a shift in the interference pattern , the time of measurement is restricted to half a cycle. Since the flux of gravitational wave energy on earth is 
extremely weak , the
common method has the disadvantage of getting extremely weak effect during this time of measurement. Another method in which the time of measurement is very long relative to the time period of the
gravitational wave can be made by the measurement of the Fourier transform of the second order correlation function. The normalized light intensity-intensity correlation function$^{20,25}$ is given
by
\begin{equation}
g^{\left( 2\right) }\left( \tau \right) =\frac{\left\langle :I\left(t\right) I\left( t+\tau \right) :\right\rangle }
{\left\langle I\left(t\right) \right\rangle ^{2}}
\end{equation}
where
I is the light intensity and : : denotes normal ordering of the electric field operators , which are in the numerator of $g^{\left( 2\right)}\left( \tau \right) $. 
The normalized light intensity-intensity
correlation function can be obtained by using 50-50 percent beam splitter , at the interferometer output , and converting the interferometer light into currents $i\left( t\right) $ and $i\left( t+\tau
\right) $ by two photomultipliers. By using a spectrum analyzer and performing the Fourier transform of the current fluctuations one gets$^{20,25-27}$:
\begin{equation}
\int\limits_{-\infty }^{\infty }\left(
\left\langle i\left( t\right) i \left(t+\tau \right) \right\rangle -\left\langle i\right\rangle ^{2}\right)e^{-i\omega \tau }
d\tau =2eG\left\langle i\right\rangle +\left\langle i\right\rangle^{2}
\int\limits_{-\infty }^{\infty }\left( g^{\left( 2\right)}\left( \tau \right) -1\right) e^{- i\omega \tau }d\tau\end{equation}
Here G is the amplification factor of the photomultiplier and in
deriving Eq. (8) a current $\delta \left( t\right) $ pulse function is assumed.$^{20,25-27}$ The first term on the right side of Eq. (8) which is known as the shot noise$^{20-21,25-27}$ can be reduced
relative to the signal by using high intensity lasers. However , other noise terms (e.g. thermal noise,seismic etc.) will affect the detection of the gravitational waves. Since we are treating the detection
of gravitational waves in the frequency range 10-10$^{4}$ Hz , there is a very strict requirement on the stability of the laser which should be coherent during a time $\tau $ which is of order the time
period of the gravitational wave $\left( 10^{-1}-10^{-4}\right) \ ,\sec $. This is the main difficulty in using the present method but with the present advanced technology this requirement can be
achieved. On the other hand we find that in using Eq. (8) one can average all terms over very long times , relative to the time period of the gravitational wave. By using Eq.(8) with the 
heterodyne term of
Eq. (6 ) one may observe certain peaks in the light intensity spectrum at frequencies $\omega _{vib\pm }\omega _{grav}$, which will indicate the observation of the periodic gravitational
wave.

\end{document}